\documentstyle[12pt]{article}

\newcommand{\bmat}{\left(\begin{array}}
\newcommand{\emat}{\end{array}\right)}
\def\NPB#1#2#3{Nucl. Phys. B{#1} (19#2) #3}
\def\PLB#1#2#3{Phys. Lett. B{#1} (19#2) #3}

\def\PRD#1#2#3{Phys. Rev. D{#1} (19#2) #3}

\def\yzero{\smash{\hbox{$y\kern-4pt\raise1pt\hbox{${}^\circ$}$}}}

\def\beq{\begin{equation}}
\def\eeq{\end{equation}}
\def\beqa{\begin{eqnarray}}
\def\eeqa{\end{eqnarray}}

\def\-{\hphantom{-}}

\def\s2{\frac{1}{\sqrt2}}

\def\beq{\begin{equation}}
\def\eeq{\end{equation}}
\def\beqa{\begin{eqnarray}}
\def\eeqa{\end{eqnarray}}

\def\IF{\relax{\rm I\kern-.18em F}}
\def\II{\relax{\rm I\kern-.18em I}}
\def\IP{\relax{\rm I\kern-.18em P}}
\def\IC{\relax\hbox{\kern.25em$\inbar\kern-.3em{\rm C}$}}
\def\IR{\relax{\rm I\kern-.18em R}}

\def\Dsl{\,\raise.15ex\hbox{/}\mkern-13.5mu D} 
\def\IZ{Z\kern-.4em  Z}

\catcode`\@=11
\newdimen\@rotdimen
\newbox\@rotbox

\def\@vspec#1{\special{ps:#1}}
\def\@rotstart#1{\@vspec{gsave currentpoint currentpoint translate
   #1 neg exch neg exch translate}}
\def\@rotfinish{\@vspec{currentpoint grestore moveto}}
\def\@rotr#1{\@rotdimen=\ht#1\advance\@rotdimen by\dp#1%
   \hbox to\@rotdimen{\hskip\ht#1\vbox to\wd#1{\@rotstart{90 rotate}%
   \box#1\vss}\hss}\@rotfinish}
\def\@rotl#1{\@rotdimen=\ht#1\advance\@rotdimen by\dp#1%
   \hbox to\@rotdimen{\vbox to\wd#1{\vskip\wd#1\@rotstart{270 rotate}%
   \box#1\vss}\hss}\@rotfinish}%
\def\@rotu#1{\@rotdimen=\ht#1\advance\@rotdimen by\dp#1%
   \hbox to\wd#1{\hskip\wd#1\vbox to\@rotdimen{\vskip\@rotdimen
   \@rotstart{-1 dup scale}\box#1\vss}\hss}\@rotfinish}%
\def\@rotf#1{\hbox to\wd#1{\hskip\wd#1\@rotstart{-1 1 scale}%
   \box#1\hss}\@rotfinish}%
\def\rotate{\@ifnextchar[{\@rotate}{\@rotate[l]}}
\def\@rotate[#1]#2{\setbox\@rotbox=\hbox{#2}\@nameuse{@rot#1}\@rotbox}

\catcode`\@=12

\topmargin
-1.5cm
\textwidth
15.5cm
\textheight
23.5cm
\oddsidemargin
0.7cm
\evensidemargin
1.2cm

\begin{document}

\makeatletter
\@addtoreset{equation}{section}
\makeatother
\renewcommand{\theequation}{\thesection.\arabic{equation}}
\pagestyle{empty}
\rightline{FTUAM-99/4; IFT-UAM/CSIC-99-4; DAMTP-1999-91}
\rightline{\tt hep-ph/9908305}
\vspace{1.5cm}
\begin{center}
\LARGE{\bf
Anomalous $U(1)$'s and Proton Stability in
Brane Models  \\[10mm]}
\large{
L.~E.~Ib\'a\~nez$^*$  and F. Quevedo$^{**}$
\\[2mm]}
\small{* \
 Departamento de F\'{\i}sica Te\'orica C-XI
and Instituto de F\'{\i}sica Te\'orica  C-XVI,\\[-0.3em]
Universidad Aut\'onoma de Madrid,
Cantoblanco, 28049 Madrid, Spain.
\\[9mm]}
\small{** \
D.A.M.T.P., Silver Street, Cambridge, CB3 9EW, UK.
\\[9mm]}

\small{\bf Abstract} \\[7mm]
\end{center}

\begin{center}
\begin{minipage}[h]{14.0cm}

We consider the most general generation-independent
$U(1)$ gauge  symmetry consistent with the presence of
 Yukawa couplings for all quarks and
leptons in the SUSY version of the Standard Model.
This $U(1)$ has  generically mixed anomalies
with SM groups,  which {\it cannot}
be cancelled by the Green-Schwarz mechanism
 of heterotic $D=4$ strings.
We argue that these anomalies can in principle  be cancelled
by the generalized Green-Schwarz mechanism
present in field theories corresponding to $D$-branes
at singularities. Moreover, unlike the heterotic case,
once the $U(1)$ symmetry is broken it may remain  as an exact
perturbative {\it global} symmetry in the low energy theory.
Applying this  scheme to the  SUSY SM we find that
gauging such a general $U(1)$:
1) $B$ and $L$ violating operators at least up
to  dim=3,4,5,6  are generically forbidden   ; 2)
The $\mu$-term is generically supressed.
We also study the properties of a $U(1)_X$ symmetry whose
 mixed anomalies with the
different SM gauge groups are in the ratio of the beta function
coefficients
 $\beta _a$.
This relation has
 been shown to hold in
certain orientifold models.
In all cases the $U(1)$ remains as a global symmetry at the
orientifold singularity, the SM Higgs can break it at the
electroweak
 scale,  making possible to relate the blowing-up of the
singularity with electroweak symmetry breaking.

\end{minipage}
\end{center}
\newpage
\setcounter{page}{1}
\pagestyle{plain}
\renewcommand{\thefootnote}{\arabic{footnote}}
\setcounter{footnote}{0}

\section{Introduction}

The SUSY standard model (SM) has a  number of naturality problems.
One of the most pressing ones is the problem of
proton stability. Indeed, the most general superpotential consistent
with $SU(3)\times SU(2)\times U(1)$ gauge invariance and
supersymmetry has dimension-three and -four operators which violate
lepton and/or baryon number. In particular it has the general form:
\beqa
 W& =& h_u\, Q_Lu_L^c\bar H + h_d\, Q_Ld_L^c H+h_l\, L_L e_L^c H
\nonumber \\
& +&  h_B\, u_L^c d_L^c d_L^c +h_L\, Q_Ld_L^cL+h'_L\, L_LL_Le_L^c
\nonumber \\
& +& \mu_L L{\bar H}\, +\  \mu H{\bar H}
\label{super}
\eeqa
in a self-explanatory notation.  The first line contains the usual Yukawa
couplings which are needed for the standard quark and lepton masses whereas
the second line shows  B or L-violating couplings and the third shows the
$\mu $-terms.
One needs to invoke a symmetry of
some sort  to forbid at least a subset of the dangerous couplings
in order to obtain consistency with proton stability.

Another problematic point of the SUSY SM is the smallness of the
$\mu $-parameter, the SUSY mass of the Higgs multiplets. In principle that
mass parameter would be expected to be as high as the cut-off of the
theory, unless there is some symmetry reason which protects the
Higgs mass from becoming large.

In field theories we are free to impose either a discrete symmetry
such as $R$-parity or global continuous symmetries to forbid the
 dangerous couplings. In string theory $R$-parity does not in general  
appear as a natural symmetry and furthermore, global symmetries
are believed not to be present. In fact, in perturbative string theory
it can be shown \cite{bd}\ that there are no global symmetries (besides
Peccei-Quinn symmetries of axion fields or accidental symmetries
of the low-energy effective action). In nonperturbative string
theory,
even though there is no general proof, it is also believed that global
symmetries are absent, the reason being that any theory that includes
gravity
will not preserve global symmetries since they are broken naturally by
black holes and other similar objects.

Therefore, perhaps the  simplest possibility for solving both problems 
in string theory would be to gauge some continuos
$U(1)$ which forbids the dangerous  couplings \cite{rabi,fiq,ir}.
 This is in general problematic
because, if we stick to the particle content of the MSSM such symmetries
are bound to be anomalous. One might think of using the
Green-Schwarz mechanism found in perturbative heterotic vacua
in order to cancel those anomalies and indeed this possibility has
been explored in the past. However there are two main obstructions:

1) The mixed anomalies of the $U(1)$'s with the SM interactions
are not in the appropriate ratios to be cancelled \footnote{ This can be
avoided by going to generation-dependent $U(1)$ symmetries
as in ref. \cite{ir2,br}. We will concentrate in this article on
flavour-independent $U(1)$ symmetries.}.

2) Due to the presence of a Fayet-Iliopoulos (FI) term, the
$U(1)$ symmetry is in general broken slightly below the
string scale and does not survive as an exact global
symmetry. Thus in general one has to rely on the particularities of
the model and holomorphicity in order to supress sufficiently
parameters like the $\mu $-term \cite{br,ramond}.

In the present letter we point out that these two problems are
not present in the alternative generalized Green-Schwarz anomaly
cancellation mechanism recently found in Type I and Type IIB
$D=4$, $N=1$ string vacua. Indeed in this novel mechanism
the mixed anomalies of a $U(1)$ with the different group
factors can be different. In addition, unlike what happens in
the perturbative heterotic case, the FI term may be put to
zero. In that case, as first pointed out  in ref.
\cite{biq},  the $U(1)$ survives as an effective
global symmetry which is exact in perturbation theory, evading in this
 way the general argument against global symmetries in string theory.
Both these aspects are wellcome  in order to
suppress the dangerous couplings with a gauged Abelian symmetry.

We will discuss two general classes of flavour-independent
 anomalous $U(1)$'s. The first
class
allows all Yukawa interactions and it is discussed in section
3 whereas another class of anomalous $U(1)$'s with anomalies
proportional to the beta function coefficients of the corresponding
gauge groups is discussed in section 4. We will start in section 2
 discussing the general aspects of anomalous $U(1)$'s in
heterotic and type I models respectively.

\section{Anomalous $U(1)$'s: Heterotic vs Type I Models}

In $D=4$, $N=1$ perturbative
heterotic vacua one anomalous $U(1)$ is often present
and anomalies are cancelled by a Green-Schwarz mechanism.
An important role is played
by the imaginary part of the complex heterotic dilaton $S$, which
is dual to the unique
antisymmetric tensor $B_{\mu\nu}$ of  perturbative  heterotic strings.
Under an anomalous $U(1)$ gauge transformation $A_\mu\rightarrow A_\mu
+\delta_{GS}\,\partial_\mu\theta(x)$,  Im$S$ gets shifted by ${\rm
Im}S\rightarrow
{\rm Im} S-\delta_{GS}\, \theta (x)$, where $\delta_{GS}$
 is a constant anomaly-cancelling coefficient.
Since the gauge kinetic function for the gauge group $G_a$  is at tree-level
$f_a=k_a\, S,$ the Lagrangian contains the couplings
${\rm Im} S \sum_a k_a F_a\wedge F_a$, where the sum runs over all
 gauge groups in
the model and the coefficients $k_a$ are known as the Kac-Moody
levels.
Then, a shift in ${\rm Im}S$ can in principle cancel mixed
$U(1)$-gauge anomalies. However, for this to be possible the
mixed anomalies $C_a$  have to be in the same ratios
\cite{ib} as the coefficients
$k_a$ (Kac-Moody levels) of the gauge factors:
\beq
{ { C_a}\over {C_b} }\ = \ { { k_a}\over {k_b} }
\label{heter}
\eeq
With $\delta_{GS}=C_a/k_a$.
 In the SM, assuming the standard hypercharge normalization,
we have $k_3:k_2:k_1=1:1:5/3$.

The  anomalous $U(1)$ induces,
 a Fayet-Iliopoulos (FI)  term
 $\xi=g^2 M_P ^2\, \delta_{GS} /16\pi
 ^2$ at one-loop
 \cite{dsw}.
The gauge coupling $g$ here is given by  $8\pi/g^2={\rm{Re}}\, S$.
The $D$-term in the Lagrangian then takes the form:
\beq
{\cal L_D} =\, \frac{g^2}{2}\,\left( \sum_i q_i X_i
K_i+\xi\right)^2
\eeq
where $K_i$ is the derivative of the K\"ahler potential $K$ with respect
to the matter fields $X_i$ which have  charge $q_i$ under the anomalous
$U(1)$. This term triggers  gauge symmetry breaking.
In order to preserve supersymmetry, the $D$-term has to vanish. The
Fayet-Iliopoulos term $\xi$ cannot vanish because, $U(1)$ being anomalous
implies
$\delta_{GS} \neq 0$ and,
in a nontrivial vacuum, $g\neq 0$. Therefore a combination
 of the charged fields
$X_i$ is forced to get a nonvanishing vev to compensate the FI term,
 breaking the anomalous
$U(1)$ and often some other non-anomalous groups.

Let us see how the anomalous $U(1)$ gauge field gets a  mass.
First, the anomaly cancelling term in the lagrangian
 $\delta_{GS}\,B\wedge F_{U(1)}$  gives rise, upon dualization, to a term
proportional to $(\partial_\mu {\rm{Im}} S+\delta_{GS} A_\mu)^2$
 which allows the
gauge field $A_\mu$ to eat the axion ${\rm{Im}} S$ and get a mass, as in
the standard Higgs effect. A linear combination of the
 real part of $S$ and the charged scalars $X_i$, also gets the
 same mass from the
Fayet-Iliopoulos term, after expanding the $D$ term around the
nontrivial vacuum. Therefore  the analogy with the supersymmetric Higgs
effect is complete: the original vector superfield eats the chiral
superfield
of the dilaton giving rise to a massive vector superfield. The
anomalous $U(1)$ symmetry gets broken at a scale determined by
$\xi$, which can be $1$-$2$ orders of magnitude below the Planck mass
depending on the value of $\delta_{GS}$.
The massless combination of the dilaton with the charged fields $X_i$
plays the role of the string coupling.

Since chiral fields $X_i$ charged under the
$U(1)$ are forced to get vevs, non-renormalizable couplings
of the form $\psi ^n X_i^m$, where the $\psi $ denote SM
superfields will induce effective operators which
will in general violate the anomalous $U(1)$ symmetry.
Thus typically this symmetry {\it does not survive}
at low energies as a global symmetry.

Let us now see how things change in Type I strings.
Recently it has been realized \cite{iru}
that the cancellation of
$U(1)$ anomalies in certain Type I and Type IIB
$D=4$, $N=1$ models
\cite{bl,ang,3kakus,zwart,odri,fin,afiv,2kakus,lpt}
 proceeds in a manner quite different to the
one in perturbative heterotic vacua.
These are models which may be constructed
as Type IIB orbifolds or orientifolds
\cite{orient} and
contain different D-brane configurations
in the vacuum. For example, the vacuum may contain
a certain number of D3-branes and D7-branes with
their transverse coordinates located at different
positions in the extra six compact dimensions.
Chiral $N=1$ theories in $D=4$ are only obtained
when e.g., the D3-branes are located on top of
orbifold singularities.
 It has been found \cite{afiv,iru,iru2}
that in this class of theories:
a) There may be more than one anomalous $U(1)$;
b) The mixed anomaly of the $U(1)$ with other groups
need not be universal ; c) There is a generalized
Green-Schwarz mechanism at work in which the
cancelling role is played not by the complex
dilaton $S$ but by twisted closed string
 massless modes $M_k$. These are fields which live
on the fixed points of the orbifold. In particular,
their real part (which are NS-NS type of fields)
parametrize the smoothing out of the orbifold
singularities whereas their imaginary parts
(which are Ramond-Ramond fields) are the ones
actually participating in the $U(1)$ anomaly
cancellation \footnote{
More precisely, from string theory, the blowing-up modes together with
the antisymmetric tensors coming from the RR sector, belong to  linear
multiplets. The scalar components of these multiplets,  $m_k$
are the ones that vanish  at the orbifold
point and their value determine the blowing-up procedure
\cite{abd}. Upon
dualization, the linear multiplets get switched to the chiral
multiplets $M_k$,
the relation between $m_k$ and the real part of $M_k$ depends on the
structure of the Lagrangian but close to the singularity it is linear
and $m_k={\rm Re}M_k - F(T_i,T_i^*)$ where  $F(T,T^*)$ is
a model dependent function which depends on the untwisted
moduli fields $T_i$ which determine the size of the
compact space.}.
 More specifically, cancellation of
$U(1)$ anomalies results from the presence in the
$D=4$, $N=1$ effective action of the term
 \beq
\sum _k \delta^l_k B_k \wedge F_{U(1)_l}
\label{shiftu}
\eeq
where $k$ runs over the different twisted sectors of the
underlying orbifold (see ref.\cite{iru,iru2} for details) and
$B^k$ are the two-forms which are  dual to the imaginary
part of the twisted fields $M_k$. Here {\it l} labels the different
anomalous $U(1)$'s and $\delta^l_k$ are model-dependent constant
coefficients. In addition the gauge kinetic functions have
also a (tree-level) $M_k$-dependent piece
\footnote{This is for gauge groups coupling either to
Type I 9-branes or 3-branes. In the case of 5-branes or
7-branes the complex field $S$ is to be replaced by
the appropriate $T_i$ field. The different choices
for Dp-branes are in fact T-dual to each other
and , hence, equivalent.  See e.g. ref. \cite{imr} for details.}
\beq
f_{\alpha }\ = \ S  \ + \ \sum _k s^k_{\alpha } M_k
\label{correc}
\eeq
where the $s^k_{\alpha }$ are model dependent
coefficients. Under a $U(1)_l$ transformation the $M_k$
fields transform non-linearly
\beq
{\rm Im} M_k \ \rightarrow {\rm Im}  M_k \ +\ \delta^l_k\Lambda _l(x)
\label{shift}
\eeq
 This non-linear transformation combined with eq.(\ref{correc})
results in the cancellation of the $U(1)_l$ anomalies
as long as the coefficients $C^l_{\alpha }$
 of the mixed $U(1)_l$-$G_{\alpha }^2$
anomalies are given by
\beq
C^l_{\alpha }\ =\ -\ \sum _k  s^k_{\alpha } \delta^l_k
\label{lares}
\eeq
Unlike the perturbative heterotic case, eq.(\ref{lares})
does not in general require universal mixed anomalies
as in eq.(\ref{heter}).

In $D=4$  models like these there can also be
mixed $U(1)_X$-gravitational anomalies. In the
perturbative {\it heterotic} case, in order
for the Green-Schwarz mechanism  to work,  the coefficient
$C^l_{grav}$ of those anomalies must be  related to those
of mixed $U(1)$-gauge anomalies by
\beq
C^l_{grav}\ =\  {{24}\over {k_{\alpha }}} C^l_{\alpha } \ .
\label{hetergrav}
\eeq
Such  relationship disappears in the case of Type IIB
$D=4$, $N=1$ orientifolds. One can find though
certain sum rules relating the gravitational to the
gauge anomalies in certain classes of models.
In particular, for the toroidal orientifolds of the
general class studied in refs.\cite{afiv} one can obtain
the constraint \cite{iru} :
\beq
C^l_{grav}\ =\ {3\over 2} \sum_{\alpha } n_{\alpha} C^l_{\alpha }
\label{orientgrav}
\eeq
where $n_{\alpha}$ is the rank of the $U(n)$ or $SO(m)$ groups
which are present in this class of orbifolds. This constraint
has to be satisfied for the anomalies to be cancelled by
the generalized Green-Schwarz mechanism present in these
models
\footnote{
It is amusing that eq.(\ref{orientgrav}), valid for certain
classes of Type IIB orientifolds, turns out to be consistent
with what is found in perturbative heterotic $SO(32)$
{\it Abelian } orbifolds. Indeed, in that case all mixed
$U(1)$-gauge anomalies are equal and one has
$\sum_{\alpha }n_{\alpha} C^l_{\alpha }= rank(SO(32))C^l$
$=16C^l$. Plugging this back into eq.(\ref{orientgrav}) we
recover the perturbative heterotic result eq.(\ref{hetergrav}).}
.

The FI term for this $U(1)_X$ is given by $\xi=\delta_X  K_M$,
since the K\"ahler potential (to first order in $M$)is given
\cite{dm,pop,iru2,abd}  by
$K=[{\rm{Re}} M - F(T_i,T_i^*)]^2+\cdots$ we can
easily see that the FI term vanishes at the orbifold singularity
\cite{afiv,iru,iru2,abd} (see also \cite{celw,lln}).
This is impossible in the heterotic case, because in that
case it is the field $S$ that cancels the anomaly and the FI-term
$\xi$ is then proportional to  $K_S\sim g^2$ which cannot be set to
zero in a nontrivial vacuum.
The anomalous gauge field gets a mass
\cite{pop,lln} exactly in the same
way as in the heterotic case, nevertheless,
 there is no need for a charged field to
get a nonvanishing vev in order to cancel the $D$-term,
  therefore the corresponding
 global symmetry is {\it not} broken.
Thus, the gauge field gets a mass of the order of the string scale
(since the mass depends on $K_{MM^*}$ and not on $K_M$) but
the global symmetry remains perturbatively exact as long as we
are at the orbifold singularity $\xi=0$. If there is a vacuum for which the 
$D$-term vanishes outside the singularity, the scale of breaking of
the global symmetry could in principle be as small as we want.

We can see then that in the class of Type IIB
orientifolds in which this anomaly cancellation mechanism
has been studied, the anomalous  $U(1)$'s have generically a
mass of order the string scale.  Unlike what happens in the heterotic
case, this FI-term is arbitrary at the perturbative level
and hence may  in principle vanish (orbifold limit).
In this case the {\it  $U(1)_X$ symmetry remains as an
effective global $U(1)$ symmetry which is perturbatively exact}.

\section{Anomalous U(1)'s and Yukawa Couplings}

We will now study the new possibilities offered by  these generalized
Green-Schwarz mechanism when applied to MSSM physics.
We will consider the simplest possibility in which we extend
the SM gauge group by adding a single anomalous $U(1)_X$.
There are just three
$U(1)$ charge asignements (beyond hypercharge)
for the MSSM chiral fields
which 1) allow for the presence of the usual Yukawa couplings
and 2) are flavour-independent. These were named $R$, $A$ and $L$
in ref. \cite{ir}, and the corresponding assignments are displayed in 
 table 1, where we also include the  hypercharge
 assignments $Y$.
\begin{table}[htb]
\renewcommand{\arraystretch}{1.25}
\begin{center}
\begin{tabular}{|c|c|c|c|c|c|c|c|}
\hline
    &  Q &  u &  d & L & e & H & ${\bar H}$ \\ \hline
6Y & -1 & 4 & -2 & 3 & -6 & 3 & -3 \\ \hline
 R & 0 & -1 & 1 & 0  & 1 & -1 & 1 \\ \hline
A & 0 & 0  & -1 & -1  & 0 & 1 & 0 \\  \hline
L & 0 &  0 & 0 & -1  & 1 & 0  & 0 \\  \hline
$Q_X$ & 0 & $-m$ & $m-n$ & $-n-p$  & $m+p$ & $-m+n$ & $m$ \\
\hline
\end{tabular}
\end{center}
\caption{$U(1)$ symmetries of the SUSY standard model.}
\label{tzn}
\end{table}
Notice that $L$ is just lepton number and  $R$
corresponds to the  3rd component of right-handed  isospin
in left-right symmetric models (baryon number is given by the
combination
$B=6Y+3R+3L$). The other symmetry, $A$,
is a Peccei-Quinn type of symmetry.
 Thus the more general such $U(1)$ symmetry will be a
linear combination:
\beq
Q_X\ = \ mR\ + \ nA\ +\ pL
\label{qex}
\eeq
where $m,n,p$ are real constants. We will denote the corresponding
$U(1)_X$'s by giving the three numbers $Q_X=(m,n,p)$. The fifth line in the
table shows the general $Q_X$ charge of the particles of the MSSM.
Notice that, depending on the values of $m,n,p$ some of the terms in the
superpotential (1.1) may be forbiden. Thus, for example, $Q_X=R=(1,0,0)$
forbids all the terms in the second line and would forbid proton
decay at this level
by itself. On the other hand it does not provide an explanation for the
smallness of the $\mu $-term. In particular, the bilinear $H{\bar H}$
has $Q_X$ charge equal to $n$, and hence it can only be forbiden if
our $U(1)$ has $n\not= 0$.
 For that purpose we can see that  the $U(1)$
symmetry
is necessarily anomalous. Thus let us study the
anomalies of the above $Q_X$ symmetry.

The mixed anomalies $C_i$  of $Q_X$ with the SM gauge interactions
are given by\footnote{We will not consider constraints coming from
cancellation of mixed $U(1)$-gravitational
anomalies in our analysis in this chapter and the next, since
one can always cancel those by the addition of apropriate SM
singlets carrying  $U(1)_X$ charges.}:
\beqa
\label{anomaly}
C_3\ & = & \ -n {{N_g}\over 2}   \\ \nonumber
C_2\ & = & \ -n {{N_g}\over 2}\ +\ n{{N_D}\over 2}\ -\ p{{N_g}\over 2} \\
\nonumber
C_1\ & = & \ -n {{5N_g}\over 6}\ +\ n{{N_D}\over 2}\ +\ p{{N_g}\over 2}
\eeqa
where we have preferred to leave the number of generations $N_g$ and  
doublets
$N_D$ free to trace the origin of the numerical factors (one has
$N_g=3$, $N_D=1$ in the MSSM). There is an additional constraint from the
cancellation of $U(1)_Y\times U(1)_X^2$ anomalies which yields:
\beq
p\ (m-n)\ = \ {n\over {2N_g} }\ \left(   N_D\ (n-2m)+2N_g\ m \right)    \  .
\label{quadr}
\eeq
Thus there are only two independent parameters out of the three $m,n,p$
if we imposse this latest constraint.

It is now easy to see that (\ref{anomaly}) cannot be satisfied in the
{\it heterotic} case since there is no solution to these equations with
 $C_3:C_2:C_1=1:1:5/3$.
 However these equations
 can, in principle, be easily satisfied in the type I case.
To see this, let us suppose for simplicity that one twisted field $M$
is relevant in the cancellation mechanism \footnote{Notice that $M$
may also denote a linear combination of several twisted fields living
at the singularity.}. The gauge kinetic functions for the SM
interactions
will have the form:
\beq
\label{gkf}
f_\alpha=S+s_\alpha\,M\,, \qquad \alpha=3,2,1.
\eeq
Now, from eq.(\ref{lares}) we see that
 the mixed anomalies will cancel if the anomaly coefficients in
(\ref{anomaly}) are related to the parameters  $\delta_X$ and
$s_\alpha$ by
 $C_\alpha =-\delta_X\times s_\alpha$, which is possible to satisfy
for appropriate $\delta_X$ and $s_\alpha$.
Notice , however, that in the present case
the coefficients $s_{\alpha }$ are related. Indeed,
for the physical case $N_g=3, N_D=1$ anomalies can be cancelled as
long
as the parameters $s_\alpha$ satisfy:
\beq
2 s_3\ =\ s_1+s_2
\label{cons}
\eeq
 This imposes
a constraint on which type I models
can have an anomalous $U(1)$  that allow all
standard Yukawa couplings.

\begin{table}[htb]\footnotesize
\renewcommand{\arraystretch}{1.25}
\begin{center}
\begin{tabular}{|c|c|c|c|c|c|c|c|c|}
\hline
    &  $Qu{\bar H}$ & $Qd{\bar H}$  &  $LeH$ &
 $H{\bar H}$ & $L{\bar H}$ & $udd$  & $QdL$  & $LLe$ \\ \hline
$Q_X$ & 0 & 0& 0 & $n$ & $m-n-p$ & $m-2n$ & $m-2n-p$ & $m-2n -p$ \\ \hline
 $Q_X$' & $-Y$ & $Y$ & $Y-3\delta_X$ & $-\delta_X$  & $-2\delta_X +Z-Y$ &
 $Z$ & $Z-\delta_X$ & $Z-4\delta_X$ \\
\hline
\end{tabular}
\end{center}
\caption{$U(1)_X$ charge of dim=3,4 operators in the MSSM
for the two classes of $U(1)$'s considered in sections
3 and 4 respectively.}
\label{tzndos}
\end{table}

Let us now be a bit more specific and study the posibilities for
anomalous $Q_X=(m,n,p)$ symmetries. As we said, we need
$n\not= 0$ in order to forbid the $\mu $-term.

\begin{enumerate}

\item{}
The simplest case
is obtained for $p=0$, i.e., no gauging of lepton symmetry.
In this case condition (\ref{quadr}) requires
$n=2m(N_D-N_g)/N_D=-4m$ and we are left  with a unique
possibility  $Q_{\mu }$ consistent with anomaly cancellation:
\beq
Q_{\mu }\ =\ R\ -4A
\label{cojo}
\eeq
It is easy to check (see Table 2)
that this symmetry forbids all dimension 3 and 4
terms violating baryon or lepton number in eq.(\ref{super}).
Dangerous $F$-term operators of dimension smaller than $9$ are also
forbidden
in this case (see for instance reference \cite{benaklidav}, for a recent
discussion of these operators). The dimension 6 operators
$[QQu^*e^*]_D$ and $[Qu^*d^*L]_D$ are however allowed
(see Table 3).

\item{}
A related symmetry is the one introduced by Weinberg in 1982
in order  to
eliminate dangerous $B,L$ violating operators
in the supersymmetric SM
\cite{wein}. In his model
all quarks and leptons carry unit charge whereas the Higgses have
charge $-2$. This symmetry corresponds to $Q_W\ =\ -5 R\ -4 A\ -6Y$.
In order to cancel $U(1)$ anomalies he was forced to add
extra chiral fields transforming like
$(8,1,0,-2)$+$(1,3,0,-2)$+$2(1,1,1,-2)$+$2(1,1,-1,-2)$
under $SU(3)\times SU(2)\times U(1)_Y\times U(1)_X$.
We now see that in the present context the addition
of all those extra fields is not required and one
can stick to the particle content of the MSSM
as long as the anomaly cancelation mechanism here
considered is at work. The $U(1)$
 clearly satisfies equations (\ref{anomaly}) with $p=0, n=-4$
but the value $m=-5$ does not satisfy the quadratic constraint
(\ref{quadr}). However this is exactly cancelled by the contribution from 
the hypercharge. Concerning what $B$, $L$-violating operators are
allowed, the same as in the previos example applies.

\item{}
For the generic case with $m,n,p\not=0$ table 2 and 3 show that all
$B$, $L$ violating operators up to dimension 6 are forbidden.
A similar analysis may be done for higher dimensional operators
which may be also dangerous for models with a relative small
string scale
\cite{lykken,untev,anton,bajogut,sundrum,shiutye,bachas,kakutye,benakli,biq,
ghilen}.

\item{}
For particular choices of $m,n,p$ one can allow some
$R$-parity violating dim=4 operator. For example,
the $U(1)$
given by:
\beq
Q_{udd}\ =\ 4R\ +\ 2A\ +\ 3L
\label{qudd}
\eeq
forbids all dim=3,4,5,6  lepton violating couplings but allow the
coupling of type  $udd$ in the superpotential.
One can also find choices
which allows for lepton number violating but not for baryon number
violating ones.

\item{}
There is a particularly simple $U(1)$ for which all the anomalies
are the same: $C_1=C_2=C_3$. This will require a string model
with identical $s_\alpha$ coefficients in the gauge kinetic function.
This is a solution as long as $N_g=3N_D$ which is satisfied in the
physical case with $p=n/3; m=-3n/2$. Since the three $s_\alpha$ are
identical
the gauge couplings are unified for any value of $<ReM>$
. Notice however that the $U(1)_Y$ normalization is not the
canonical one.
This $U(1)$ also forbids
all dangerous couplings of dimension $3,4,5$ and $6$ as well as all
dangerous
$F$-term operators of dimension less than $9$.

\end{enumerate}

\begin{table}[htb]
\renewcommand{\arraystretch}{0.75}
\begin{center}
\begin{tabular}{|c|c|c|c|}
\hline
      Operator &  Dimension &  $Q_X$ Charge & $Q_X'$ charge
 \\ \hline\hline
$[QQQL]_F$ & 5 & $-n-p$ & 0   \\ \hline
 $[uude]_F$ & 5 & $p-n$ & $-\delta_X$    \\ \hline
$[QQQH]_F$ & 5 & $n-m$  & $\delta_X+Y-Z$   \\  \hline\hline
$[\bar H\bar H e^*]_D$ & 5 &  $m-p$ & $Z-2Y$  \\  \hline
$[QuL^*]_D$ & 5 & $p+n-m$ & $2\delta_X-Z$  \\ \hline
$[\bar H^* H e]_D$ & 5 & $p+n-m$ & $2Y-\delta_X-Z$ \\ \hline
$[QQd^*]_D$ & 5 & $n-m$ & $\delta_X-Z$  \\ \hline  \hline
$[uuuee]_F$ & 6 & $2p-m$ & $-2\delta_X-Z$  \\ \hline
$[uddH\bar H]_F$ & 6 & $m-n$ & $Z-\delta_X$  \\ \hline
$[dddLH]_F$ & 6 & $2m-3n-p$ & $Y-2\delta_X+2Z$  \\ \hline
$[uddL\bar H]_F$ & 6 & $2m-3n-p$ & $2Z-2\delta_X-Y$  \\ \hline \hline
$[AA^*LH^*]_D$ & 6 & $m-2n-p$ & $Z-\delta_X-Y$  \\ \hline
$[AA^*L\bar H]_D$ & 6 & $m-n-p$ & $Z-Y-2\delta_X$  \\ \hline
$[QQQ\bar H^*]_D$ & 6 & $-m$ & $2\delta_X+Y-Z$  \\ \hline
$[QQu^*e^*]_D$ & 6 & $-p$ & $ 2\delta_X$  \\ \hline
$[Qu^*d^*H]_D$ & 6 & $2n-m$ & $Y-Z$   \\ \hline
$[Qu^*d^*L]_D$ & 6 & $-p$ & $-\delta_X$  \\ \hline
$[Qu^*d^*\bar H^*]_D$ & 6 & $n-m$ & $\delta_X+Y-Z$  \\ \hline
$[Qd^*d^*\bar H]_D$ & 6 & $2n-m$ & $-Y-Z$  \\ \hline
$[Qd^*d^*L^*]_D$ & 6 & $3n-2m+p$ & $2\delta_X-2Z$  \\ \hline
$[Qd^*d^*H^*]_D$ & 6 & $n-m$ & $\delta_X-Y-Z$  \\ \hline
$[Qu\bar H^* e]_D$ & 6 & $p-m$ & $Y-Z$  \\ \hline
$[QdH^*e^*]_D$ & 6 & $m-2n-p$ & $2\delta_X-Y+Z$  \\ \hline
$[Qd\bar H e^*]_D$ & 6 & $m-n-p$ & $\delta_X-Y+Z$  \\ \hline
$[LLH^*H^*]_D$ & 6 & $2m-4n-2p$ & $2Z-2Y-2\delta_X$  \\ \hline
$[ddde^*]_D$ & 6 & $2m-3n-p$ & $\delta_X+2Z$  \\   \hline
\end{tabular}
\end{center}
\caption{Supersymmetric operators of dimension 5 and 6 that violate
Baryon or Lepton numbers, including their charges with respect to the
anomalous $U(1)$'s of sections 3 and 4. Here the fields $A$ represent
any of the fields of the supersymmetric standard model.}
\label{tzncuatro}
\end{table}

Thus one concludes that ensuring a small $\mu$-parameter
implies in general that $B$ and $L$-violating dim=3,4,5,6  operators
are generically forbidding in the presence of anomalous
$U(1)$'s of this type except  for very particular cases.
 As for neutrino masses, the operator $LL{\bar H}{\bar H}$
is forbiden as long as $m\not=n+p$ so one must include right-handed
neutrinos to allow for neutrino masses
with charges $n+p-m$.

Now, once the $U(1)_X$ symmetry is gauged, the $\mu $
parameter is forced to be zero at the perturbative level
and we understand why the Higgs fields have small
masses compared to the cut-off. A small but
non-vanishing $\mu $-parameter is however needed
in order to obtain appropriate $SU(2)\times U(1)$
symmetry breaking. Notice however that once
SUSY  is broken,
a non-vanishing $D_X$-term will in general appear
of order $M_W^2$. Thus the
 $U(1)_X$ symmetry will get small breakings of
order $M_W$ and an effective $\mu $-term of order
$M_W$ could be generated.
$U(1)_X$ symmetry breaking effects could also be
generated from  non-perturbative effects and could
also give rise to a $\mu $ term.

\section{Anomalous $U(1)$'s and $\beta$-functions}

The class of $U(1)$ symmetries considered in the previous section is
very interesting phenomenological and in principle it
may be realized in type I
strings.
However, at the moment we
do not yet know an explict example
giving rise to such symmetries since we are still
lacking sufficiently realistic models.
We will now change our approach in the following way. Instead of
imposing the phenomenological requirements of allowing all quark and
lepton masses, we will impose some constraints on the
anomalies inspired in some known orientifold models.
Recently a special class of anomalous $U(1)$'s has been found in
orientifold models \cite{imr,iru2,ib2}.
 These models are such that the mixed anomalies
of these $U(1)$'s with the other gauge groups are in
the ratio of the beta-function coefficients of the corresponding
gauge groups. This could be of interest also in trying to
accomodate a string scale well below the unification scale
$M_X=2\times 10^{16}$ GeV \cite{imr,ib2}.
For these $U(1)$'s instead of (3.1), valid for heterotic
models, we would  have:
\beq
\frac{C_{\alpha }}{C_{\gamma }}=\frac{\beta_{\alpha }}
{\beta_{\gamma}}
\eeq
If an extension of the MSSM exists with such an extra $U(1)_X$,
since we know the beta-function coefficients for the supersymmetric
standard model, $\beta_3:\beta_2:\beta_1=-3:1:11$,
 we can then look for the most general family
independent
$U(1)$ that satisfies these constraints.
Imposing the three constraints for the mixed anomalies with the
standard model groups we find four independent solutions as shown in
the table. The most general anomalous $U(1)$ is
\beq
Q_X'\ =\ m\ Q_1+n\ Q_2+p\ Q_3+q\ Q_4
\label{qxprime}
\eeq
 which has mixed anomalies
with the standard model groups given by:
\beq
C_{\alpha }\ =\ - { {\beta _{\alpha }}\over 2} (2m+2n+p+q)
\label{cas}
\eeq
We therefore will assume $2m+2n+p+q\not=0$ so that the $U(1)$ is
indeed anomalous. These anomalies are cancelled if
the gauge kinetic function has the form $f_{\alpha }=S+{{\beta _{\alpha }}
\over 2}M$ (i.e., $s_{\alpha}=\beta _{\alpha }/2$) with
$\delta_X=(2m+2n+p+q)$. We thus have
$C_{\alpha }\ =\ - { {\beta _{\alpha }}\over 2}\delta_X$.

 \begin{table}[htb]
\renewcommand{\arraystretch}{1.25}
\begin{center}
\begin{tabular}{|c|c|c|c|c|c|c|c|}
\hline
    &  Q &  u &  d & L & e & H & ${\bar H}$ \\ \hline
 $Q_1$ & 1 & 0 & 0 & -3  & -2 & 0 & -2 \\ \hline
$Q_2$ & 1 & 0  & 0 & -3  & -2 & -2 & 0 \\  \hline
$Q_3$ & 0 &  1 & 0 & 0  & -3 & -1  & 0 \\  \hline
$Q_4$ & 0 &  0 & 1 & 0  & -2 & -1  & 0 \\  \hline
$Q_X'$ & $m+n$ & $p$ & $q$ & $-3(m+n)$  & $-(2m+2n+3p+2q)$ & $-2n-p-q$
    &
 $-2m$ \\
\hline
\end{tabular}
\end{center}
\caption{Anomalous $U(1)$ symmetries with mixed anomalies proportional
to beta function coefficients.}
\label{tzntres}
\end{table}

There are two more conditions that have to be imposed. First that the
mixed $U(1)_X-U(1)_X-Y$ anomaly vanishes identically, imposing a
quadratic constraint among the $U(1)_X$ charges. Second, that
the $U(1)_X^3$ anomaly is also cancelled by a Green-Schwarz
mechanism, therefore  a constraint $C_X=-\delta_{X}/2 \beta_x$ has
also to
be satisfied. The quadratic constraint can be automatically satisfied 
by use of the following argument: adding a term proportional to the
hypercharge
to each of the $U(1)_X$ charges will not change any of the linear
constraints
coming from the mixed $U(1)_X-G-G$ anomalies because we know that
hyercharge is anomaly free. Therefore this will change only the
quadratic
constraint, we then assume that the proportionality constant has been
fixed and not impose the quadratic constraint. This will tell us
though that the four coefficients $m,n,p,q$ are not independent and
we are free to impose one constraint among them (the Weinberg $U(1)$
of the previous section was obtained in this way). As for the cubic
constraint, since it involves only the anomalous $U(1)_X$ we will
allow the possibility of extra matter fields charged only under the
anomalous $U(1)_X$ but not the Standard Model groups, which is very
common
in string models, so that their charges satisfy the cubic anomaly condition.

Let us then try to extract the possible implications of the $U(1)_X$
symmetry. It turns out that one can draw some general conclussions
without needing to go into the details of each $U(1)_X$.
 We show in table 2 the $U(1)_X$
charges of the dim=3,4 operators in the MSSM.
Here
one defines $Y=m-n-p$ and $Z=p+2q$
\footnote{Note that $m,n,p,q$ are defined in eq.(
\ref{qxprime}) and have nothing to do with
those defined in
eq.(\ref{qex}).}. For any of those couplings to
be allowed the corresponding entry has to vanish. Examining the
table one reaches the following conclussions:

\begin{enumerate}

\item{}The $\mu$-term is prohibited as long as the
$U(1)$ is anomalous ($\delta_X\not=0$). This result happens to   be
 identical to the case in the
previous section. Thus this a very generic fact: $U(1)'s$
forbidding the $\mu$-term are necesarily anomalous.

 \item{}If a mass term for the up quarks is allowed ($Y=0$), then
automatically
a mass term for the down quarks is also allowed. However,
 at the same time, lepton
masses
are forbidden. Therefore with this $U(1)_X$  lepton or quark mass terms
may be present
but not both simultaneously.
This implies that the class of $U(1)$'s studied in the
previous section does  not fall into the present cathegory.
In the following we will consider the case that quark masses are
permitted ($Y=0$),
 similar conclusions can be obtained if only the lepton
masses
are present, or none of them. We will comment below how
leptons could get a mass.

\item{} If the baryon number violating operator $udd$ is allowed ($Z=0$)
 then the lepton number violating operator $QdL$ is
automatically
 forbidden
implying, at this level, proton stability.

\item{} From table 3 one also observes that in the generic case all
$B$ and $L$-violating terms are forbidden at least up to dimension 6
except for the dim=5 operator $[QQQL]_F$ which is always
necesarily allowed for a $U(1)$ of this type.

\end{enumerate}

One can also check that the charges of
the dimension 5 operators $LL\bar H\bar H$
are given by $2Z-2Y-4\delta_X$. Thus for
choices with $2Z=2Y+4\delta_X$
 neutrino masses can be naturally generated as in
the standard
see-saw mechanism. Alternatively, right-handed neutrinos
may be added.

We can then see how restrictive a single anomalous symmetry can be
regarding the physically interesting couplings in the superpotential.
Even though the lepton masses are forbidden, there is a very economical
way to generate them. If there is a standard model singlet $N$ with charge
$3\,\delta_{X}$, the coupling $NLeH$ is invariant under the anomalous
$U(1)$
and if $N$ has a nonvanishing expectation value, it gives rise to
lepton masses. This looks like dangerous since, as we discussed in
the introduction, once we give large vevs to fields charged under the
anomalous $U(1)_X$, this symmetry will be broken in the effective
Lagrangian, as it generically hapened in the heterotic models.

Interestingly enough there is an unexpected unbroken
discrete $Z_3$ gauge group which saves the day.
Indeed, a
 vev for the field $N$ turns out to break $U(1)_X$ to a discrete
 $Z_3$ subgroup. This is due to the fact that
the $N$ field has to have charge $3\,\delta_{X}$ whereas
the forbidden terms would require
vevs of fields with charges $\pm 2\, \delta_X$
or $\pm 1\, \delta_X$ to be allowed.
This is enough to
forbid the dangerous couplings, such as the $\mu$-term and the $B,L$
violating operators.
The ratio $<N>/M_s$ may be at the origin of a hierarchy
of fermion masses. The only dangerous coupling that is not forbidden
by this $U(1)_X$ (nor its residual $Z_3$ once leptons get masses)
is the dimension 5 operator $QQQL$. We may hope that
an extra symmetry, possibly a flavour-dependent $U(1)$ or even
a sigma-model symmetry as those discussed in \cite{iru2},
 may be at work to forbid
this operator and keep the proton stable. Notice that this coupling
is dangerous only for the first families of quarks and leptons.
A detailed analysis of flavour-dependent anomalous $U(1)$'s may also be
interesting \cite{ir2,br,ramond},
in order to study the possible structure of fermion masses. We  hope to
report on this in a future publication.

\section{Final comments}

We have studied the possible use of anomalous $U(1)$
gauge symmetries of the class found in Type IIB
$D=4$, $N=1$ orientifolds to restrict Yukawa couplings
and operators in simple extensions of the MSSM in which
a single such $U(1)$ is added.
We have studied in detail two general classes of
flavour independent
anomalous $U(1)$'s that may come from type I strings.
The general properties of anomaly cancellation
and induced FI terms are very different compared to
those previously considered in the context
of perturbative heterotic models.
Besides its intrinsic interest, this study may lead us
to extract general properties of these models.
We have seen that in most cases, dangerous couplings are forbidden,
in particular the $\mu$ term and $B, L$ violating operators are
naturally
constrained by these anomalous symmetries.
We have checked that generic symmetries of this class
easily forbid B,L violating couplings up to large
operator dimensions. This could be wellcome
for string models with the string scale well below
the Planck mass as in refs.
\cite{lykken,untev,anton,bajogut,sundrum,shiutye,bachas,kakutye,
benakli,biq,ghilen}.

It would be interesting to extend the present analysis to the case
of flavour-dependent $U(1)$ symmetries which could restrict the
patterns of fermion mass textures. Notice that in the
analysis in section 3
we have tacitely assumed that the
residual global $U(1)_X$ symmetry is
 broken (by vevs of charged scalars)
only close to the weak scale so that proton decay
supression is sufficiently efficient and a large $\mu $-term is not
generated. This is a possibility which is not pressent in
heterotic models in which the FI-term generically forces
charged scalars to get vevs close to the
string scale. However in the Type I context
 other flavour-dependent $U(1)$'s
can be assumed to be broken by vevs of singlet scalars slightly
below the string scale so that schemes analogous to those
considered in\cite{ir2,br,ramond} are also possible.
Notice also that more than one anomalous $U(1)$ may be
present now.

 If there is more than one $M$ field at the singularity, since only
one gets swallowed by the $U(1)_X$ to become massive, they can play the role
of invisible axions and solve the strong CP-problem as proposed in ref.
\cite{biq}.
 Indeed, these other fields
would be massless to high accuracy and have the adequate couplings to
do
 the job.
It is unlikely that they get substantial masses after SUSY-breaking if
the latter originates in a hidden sector. However it is not clear if these
fields couple to $F\wedge F$. For example, in $Z_3$ with 9-branes,
the sum of
27 fields is massive, the other 26 are not. But they have zero coupling
with $F\wedge F$. We expect that in the generic case, for an
anomalous $U(1)_X$, the linear combination $s^k_X M_k$ cancels the
anomaly and gets eaten by the anomalous gauge field, whereas
the combination $s^k_3 M_k$  is the one that couples to QCD
 and plays the role of the QCD axion.

 Once $SU(2)\times U(1)$ is broken and the Higgs fields get vevs, those 
vevs will also break the $U(1)_X$  symmetry. In the $D_X$ term this can
be compensated by a tiny (compared to the string scale) FI-term $\xi$.
 Thus
it seems electroweak symmetry breaking will trigger a FI-term of
order $M_W$.
This means that $<M> \propto M_W$ in these
models.
 Thus, interestingly
enough, the electroweak scale would be a measure of the
blowing up of the singularity. The process of $SU(2)\times U(1)$
breaking would look like a transition of some branes to the bulk.
The distance of the branes to the original singularity (given by the vevs of
the Higgs) is equal to the induced FI term. Of course, all this depends
on the supersymmetry breaking mechanism and how it affects the
structure of the $D$-terms.

In the general case  we can say that
for an arbitrary anomalous $U(1)$  the vacuum is either
at the singularity $\xi=0$ or not. If it is not, then the blowing-up mode 
can substantially affect the unification scale as argued in refs.
\cite{imr,ib2}.
 Otherwise the anomalous $U(1)$ symmetry remains
as a perturbatively exact global symmetry that can help forbidding
dangerous couplings allowed by supersymmetry. This is the first
concrete proposal to  evade
 the general claim against the existence of
global symmetries in string models.
 The SM Higgs can break
this symmetry,
 triggering  a nonvanishing value to the FI term and then
moving away from the singularity. This may provide a `brany' interpretation
to the scale of electroweak symmetry breaking. In any case these new
anomalous symmetries can certainly play a very interesting role in
low-energy physics.
\bigskip

\centerline{\bf Acknowledgements}
This work has been partially supported by
 CICYT (Spain), the European Commission (grant
ERBFMRX-CT96-0045),
the John Simon Guggenheim foundation and PPARC.

\end{document}